\documentclass[reprint, amsmath, amssymb, aps, superscriptaddress]{revtex4-1}
\usepackage{packages}

\begin{document}

\preprint{APS/123-QED}

\title{Linear Response of a Periodically Driven Thermal Dipolar Gas}

\author{Reuben R.W. Wang}
\affiliation{JILA and Department of Physics, University of Colorado Boulder, Boulder, Colorado 80309, USA}
\author{Andrew G. Sykes}
\affiliation{Transpower New Zealand Limited, Wellington, New Zealand}
\author{John L. Bohn}
\affiliation{JILA and Department of Physics, University of Colorado Boulder, Boulder, Colorado 80309, USA}

\date{\today} 

\begin{abstract}

We study the nonequilibrium dynamics of an ultracold, non-degenerate dipolar gas of $^{164}$Dy atoms in a cylindrically symmetric harmonic trap. To do so, we investigate the normal modes and  linear response of the gas when driven by means of periodic modulations to the trap axial-frequency. We find that the resonant response of the gas depends strongly on the dipole alignment axis, owing to anisotropies in the differential cross section of the atoms.  We employ the use of the method of averages as well as numerical Monte Carlo methods for our analysis.  A striking result is that certain normal modes, termed ``melting modes'', initiated in an anisotropic out-of-equilibrium configuration, relax to equilibrium without oscillating.  

\end{abstract}

\maketitle

\section{\label{sec:introduction} Introduction}

The science of ultracold matter was greatly enriched with the ability to cool and trap highly magnetic species such as chromium \cite{Weinstein02_PRA,Griesmaier05_PRL,Chicireneau06_PRA}, dysprosium \cite{Newman11_PRA, Lu11_PRL, Lu12_PRL, Tang15_NJP}, and erbium \cite{Aikawa12_PRL, Aikawa14_PRL_1}.  The distinct anisotropy of the dipole-dipole interaction among atoms like these has led to a host of novel phenomena in degenerate Bose and Fermi gases.  These include such things as magnetostriction \cite{Stuhler07_JMO}, quantum Rosenzweig instability \cite{Kadau16_Nat}, self-bound dipolar droplets \cite{Schmitt16_Nat, Chomaz16_PRX}, anisotropic Fermi surface \cite{Aikawa14_Sci}, quantum Newton's cradles \cite{Tang18_PRX}, and most recently, even a supersolid phase \cite{Guo19_Nat,Chomaz19_PRX}.  

Comparatively little attention has been paid to {\it thermal} dipolar gases at ultralow temperatures, that is at temperatures above $T_c$ for bosons or well above $T_F$ for fermions, where the dynamics of the gas obeys Maxwell-Boltzmann statistics. Here the dipolar nature of the atoms can also be quite significant. At a sufficiently low temperature $T$, and in a modest magnetic field, the Zeeman splitting can exceed the mean kinetic energy $k_BT$, assuring that the atoms can remain spin-polarized in their ground states. Dynamics of the gas is then dominated by the highly anisotropic cross section of the colliding dipoles \cite{Hensler03_APB,Bohn14_PRA}.

This anisotropy is made manifest when the gas is taken out of equilibrium.  For example, when the gas is suddenly compressed in a certain direction, collisional relaxation will cause its mean kinetic energy in the transverse direction to rise, a process known as cross-dimensional rethermalization.  The rate of this rethermalization is a strong function of  the direction the dipoles are tilted with respect to the excitation axis \cite{Bohn14_PRA, Sykes15_PRA}.  This effect was first demonstrated in fermionic $^{167}$Er and readily explained using the microscopic differential cross section of the dipoles \cite{Aikawa14_PRL}.  It was subsequently extended to bosonic $^{162}$Dy and $^{164}$Dy,  and used to make the first identification of the $s$-wave scattering length of dysprosium \cite{Tang15_PRA}. Dipolar collisions also influence the aspect ratio of the gas as it expands freely \cite{Tang16_PRL}.

More broadly, ultracold but thermal dipolar gases may possess rich anisotropic dynamics when taken out of equilibrium.  In this paper we take the first steps to characterize such a gas, focusing on the regime of weak periodic drives and linear responses to emphasize the novelties inherent in dipolar scattering. In this regime, we derive a method-of-averages model that incorporates the basic physics at play and allows the determination of normal modes and their damping. We validate this model by comparing its results to those of a Monte Carlo simulation.  The results show a strong anisotropy in the response of the gas as the polarization axis is tilted with respect to the direction along which the drive is applied.  We further characterize the gas' response in terms of its normal modes, much as was done previously for a gas of hard spheres \cite{Guerey99_PRA}. This will lay the groundwork for future investigations where the direction of dipolar polarization becomes a handle with which to study, manipulate, and perhaps even exploit the anisotropic thermodynamics of the gas.

This paper is organized as follows. In Sec.~\ref{sec:BoltzmannEqn}, we provide an overview of the physical system and discuss its relevant details. In Sec.~\ref{sec:theoreticalMethods}, we briefly introduce the classical Boltzmann equation and tools employed to solve it. These tools include the method-of-averages, which allows a derivation of the Enskog equations (Sec.~\ref{sec:MethodOfAverages} and Sec.~\ref{sec:AnisotropicScattering}); and numerical Monte Carlo methods (Sec.~\ref{sec:NumericalSim}). We examine the validity of the Enskog equations in Sec.~\ref{sec:results}, then use it to investigate the normal modes of the gas and its linear response in Sec.~\ref{sec:LinearResponse}. Remarks on the exclusion of dipolar mean-field effects are provided in Sec.~\ref{sec:MeanField}, and conclusions are drawn in Sec.~\ref{sec:conclusions} with possible avenues for future works.

\section{\label{sec:BoltzmannEqn} Formulation}

\begin{SCfigure}[][b]
\centering
\pgfplotsset{colormap/hot2}
\begin{tikzpicture} 
    \coordinate (orig) at (3.42, 2.85);
    \coordinate (z) at (3.42, 3.5);
    \coordinate (epsilon) at (2.25, 3.75);
    
    \begin{axis}
        [view={135}{20},
        axis lines=center, axis on top, ticks=none,
        set layers=default, axis equal, 
        xlabel={\Large$x$}, ylabel={\Large$y$}, zlabel={\Large$z$},
        xlabel style={anchor=south east},
        ylabel style={anchor=south west},
        zlabel style={anchor=south west},
        enlargelimits, tick align=inside, domain=0:3.00, samples=20, z buffer=sort, 
        ]
        \addplot3 [surf, red, opacity=0.2, domain=-1:0,
        domain y=0:360] ({0.5*sin(y)*sqrt(1-x^2)},{0.5*cos(y)*sqrt(1-x^2)},{x});
        \addplot3 [surf, red, opacity=0.2, domain=0:1,
        domain y=0:360,on layer=axis foreground] ({0.5*sin(y)*sqrt(1-x^2)},{0.5*cos(y)*sqrt(1-x^2)},{x});
    \end{axis}
    
    \draw[->, blue, thick] (orig) -- (epsilon) node[above] {\Large$\hat{\boldsymbol{\varepsilon}}$};
    \pic [draw, ->, thick, "\Large$\boldsymbol{\alpha}$", blue, angle eccentricity=1.5] {angle = z--orig--epsilon};
\end{tikzpicture}
\caption{The initial state of the gas at thermal equilibrium in a cylindrically symmetric harmonic trap, elongated along the $z$ axis. The dipole alignment axis $\hat{\boldsymbol{\varepsilon}}$, lies in the $x$-$z$ plane, for which the dipole alignment angle $\alpha$, is the angle of inclination from the $z$ axis. }
\label{fig:dipolar_gas_init}
\end{SCfigure}
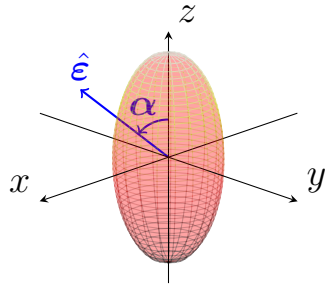

We consider a gas of $N$ magnetic atoms (dysprosium 164 in the examples  below), harmonically trapped in a cylindrically symmetric confining potential
\begin{align}
    U(\boldsymbol{q}) = \frac{1}{2} m \left[ \omega_z^2 z^2 + \omega_{\perp}^2 \left(x^2 + y^2\right) \right]. \label{eq:trap_potential}
\end{align}
For concreteness, we will take the gas to be weakly trapped along the axial direction $z$, while tightly trapped along the radial directions $x$ and $y$ (i.e. $\omega_z < \omega_{\perp}$). This identifies a unique direction in space, ${\hat z}$, with respect to which the direction of the dipoles' polarization is defined.  The geometry of the model is shown in Fig.~\ref{fig:dipolar_gas_init}, where the dipoles are assumed to be polarized in a direction $\hat{\boldsymbol{\varepsilon}}$ with respect to the $z$ axis, and make an angle $\alpha$ with respect to it. The gas is assumed to be initially in thermal equilibrium at a temperature $T_0$ which is above the critical temperature for Bose-Einstein condensation if the atoms are bosons, and well above the Fermi temperature if they are fermions.  The gas therefore obeys Maxwell-Boltzmann statistics.

We are interested in the linear response of the gas to a weak, periodic drive. This is generated by a periodic modulation of the trap frequency along the symmetry axis,
\begin{align}
    \omega_z^2(t) = \omega_{z, 0}^2 \left[ 1 + \delta \sin(\Omega t) \right], \label{eq:time_varying_trap}
\end{align}
where the drive amplitude $\delta$ is a dimensionless amplitude that is small compared to unity.

In response to the drive, the gas will warm up. Generally, the rate at which the heating occurs is governed by the differential cross section $d \sigma / d \Omega$ of the atoms.  This cross section in turn depends on three parameters of the atoms: 1) their $s$-wave scattering length, $a$ (for bosons, as we will assume here); 2) their magnetic dipole moment $\mu$, expressed as a magnetic dipole length $a_d = m\mu_0\mu^2 / (8\pi\hbar^2)$ where $\mu_0 = 1.257 \times 10^{-6}$ [H/m] is the vacuum permeability; and 3) 
the angle $\alpha$ between the direction of the dipoles' polarization axis and the symmetry axis of the trap.  The response of the gas as a function of these three parameters, as well as the frequency $\Omega$ of the drive, is our subject.

\section{\label{sec:theoreticalMethods}Theoretical Methods}

The dynamics of a gas of particle of mass $m$ is given by the evolution of its phase space distribution $f(\boldsymbol{q},\boldsymbol{p}, t)$ of coordinates $\boldsymbol{q}$ and momenta $\boldsymbol{p}$, as governed by the classical Boltzmann equation
\begin{align}
    \left[\frac{\partial }{\partial t} + \frac{1}{m}\boldsymbol{p}\cdot\grad_q + \boldsymbol{F}\cdot\grad_p \right]f = \mathcal{I}[f]. \label{eq:Boltzmann_equation}
\end{align}
Here $\boldsymbol{F} = -\grad U(\boldsymbol{q}) - \grad U_{\rm mf}(\boldsymbol{q})$ is the force applied to the atoms, which may originate both from the applied trapping potential $U$, and from the mean field interaction $U_{\rm mf}$ due to the other atoms.  We will argue below, however, that $U_{\rm mf}$ is irrelevant to our simulations. 

Energy and momentum are redistributed in the gas via two-body collisions, incorporated in the collision integral
\begin{align}
    \mathcal{I}[f] = \int d\Omega_{p'} \frac{d\sigma}{d\Omega_{p'}} \int \frac{d^3p_1}{m} \norm{\boldsymbol{p} - \boldsymbol{p}_1} \left[f'f_1' - ff_1\right]. \label{eq:collision_integral}
\end{align}
As is conventional, this expression uses the shorthand notations $f = f(\boldsymbol{q}, \boldsymbol{p}, t)$ and $f_1 \equiv f(\boldsymbol{q}_1, \boldsymbol{p}_1, t)$ for distributions of the collision partners, while primes  indicate their post-collision distributions.

For dipolar bosons at ultracold temperatures, the differential cross section was derived in \cite{Bohn14_PRA}.  As mentioned, the dipoles are assumed polarized so that their dipole moments lie along an axis $\hat{\varepsilon}$ fixed in the laboratory.  We will take this axis to lie in the $x$-$z$ plane, so that its coordinates in the laboratory frame are $\hat{\boldsymbol{\varepsilon}} = (\varepsilon \sin \alpha, 0, \varepsilon \cos \alpha)$. The differential cross section is given by $d\sigma / d\Omega (\hat{\boldsymbol{p}}, \hat{\boldsymbol{p}}^{\prime}) = |f_\mathrm{scat}(\hat{\boldsymbol{p}}, \hat{\boldsymbol{p}}^{\prime})|^2$ in terms of the scattering amplitude
\begin{widetext}
\begin{equation}
    f_{\mathrm{scat}}\left(\hat{\boldsymbol{p}}, \hat{\boldsymbol{p}}^{\prime}\right) = \frac{a_d}{\sqrt{2}}\left[-2 \left( \frac{a}{a_d} \right) - \frac{2 \left(\hat{\boldsymbol{p}} \cdot \hat{\boldsymbol{\varepsilon}} \right)^{2}+2\left(\hat{\boldsymbol{p}}^{\prime} \cdot \hat{\boldsymbol{\varepsilon}}\right)^{2}-4(\hat{\boldsymbol{p}} \cdot \hat{\boldsymbol{\varepsilon}})\left(\hat{\boldsymbol{p}}^{\prime} \cdot \hat{\boldsymbol{\varepsilon}}\right)\left(\hat{\boldsymbol{p}} \cdot \hat{\boldsymbol{p}}^{\prime}\right)}{1-\left(\hat{\boldsymbol{p}} \cdot \hat{\boldsymbol{p}}^{\prime}\right)^{2}}+\frac{4}{3}\right] \label{eq:scattering_amplitude}
\end{equation}
\end{widetext}
where $\hat{\boldsymbol{p}}$ is the unit vector denoting the pre-collision relative momentum between scatterers and $\hat{\boldsymbol{p}}'$ is the unit vector for the post-collision relative momentum.

The gas is assumed to be prepared in thermal equilibrium at time $t=0$, for which the phase-space density distribution function adopts the Maxwell-Boltzmann distribution
\begin{align}
    & f_{\text{eq}} (\boldsymbol{q}, \boldsymbol{p}) = \frac{N}{Z} \exp\left[ -\frac{p^2/2m + U(\boldsymbol{q})}{k_B T_0} \right],
    \label{eq:initial_distribution}
\end{align}
where $N$ is the system number of particles, $T_0$ is the initial temperature and $Z = \int d^3 q d^3 p \exp\left[ -\frac{p^2/2m + U(\boldsymbol{q})}{k_B T} \right]$. Starting at time $t = 0$, the trap is driven according to (\ref{eq:time_varying_trap}).

As the driven gas evolves in time, it will begin to heat, conceivably at different rates in the different directions.  To track this heating, we define a trio of pseudotemperatures by evaluating the mean kinetic and potential energies, in the three coordinates,
\begin{equation}
    \mathcal{T}_{j} = \frac{m \omega_{j}^{2} \langle {q_j}^{2} \rangle}{2 k_{B}} + \frac{\langle {p_{j}}^{2} \rangle}{2 m k_{B}}, \label{eq:pseudotemperatures}
\end{equation}
where the brackets denote an ensemble average. The evolution of these pseudotemperatures with time is one of the key observables in our results. 

In the following we will develop two alternative ways of computing the time evolution of the gas.  One is the method of averages, which simplifies the theory by tracking appropriate mean values over time, rather than the full phase space distribution.  The second is a numerically robust Monte Carlo method.  Agreement between the two methods both validates the approaches, and identities the limits of the linear response regime.

\subsection{\label{sec:MethodOfAverages} The Method of Averages}

The gas is assumed to start at thermal equilibrium, whereby its phase space distribution $f(\boldsymbol{q}, \boldsymbol{p}, t)$ is a Gaussian function of the phase space coordinates (\ref{eq:initial_distribution}).  For a weak enough driving amplitude, it is plausible that $f$ remains approximately Gaussian \cite{Krook1977}, whereby the collision integrals may be done analytically to linear order. We therefore present analytic \textit{linearized} solutions to these 2-body collision  integrals similar to that done in \cite{Guerey99_PRA, Colussi15_NJP}, but for anisotropic differential cross sections. To enact this linearized {\it ansatz},  $f$ is assumed to be a Gaussian function at all times, characterized by the  time-dependence of the spatial and momentum variances $\langle q_j^2 \rangle$ and $\langle p_j^2 \rangle$ \footnote{The symmetry of the trap allows us to assert that $\langle q_j \rangle = \langle p_j \rangle = 0$ (i.e. no center of mass motion).}.

The linearized approximation thus admits the equations of motion for the mean values of dynamical variables $\chi(\boldsymbol{q}, \boldsymbol{p})$. Such an approach is known as the method of averages \cite{Reif} and is performed by defining the phase space averages
\begin{align}
    & \langle \chi \rangle \equiv \frac{1}{N} \iint d^3p d^3 q \: f(\boldsymbol{q}, \boldsymbol{p}, t) \chi (\boldsymbol{q}, \boldsymbol{p}, t),
\end{align}
where
\begin{align}
    & N = \iint d^3p d^3 q f(\boldsymbol{q}, \boldsymbol{p}, t).
\end{align}

The equations of motion governing $\langle \chi \rangle$ can then be derived by multiplying the Boltzmann equation by $\chi$ and integrating over all of phase-space, 
\begin{align}
    \frac{1}{N} \int d^3p d^3 q \: \chi D f = \frac{1}{N} \int d^3p d^3 q \: \chi \mathcal{I}[f],
\end{align}
with $D$ being the substantial derivative. To derive a self-consistent set of equations of motion, we require $\chi$ to come from the set of nine variables $\{q_j^2, p_j^2, q_j p_j\}$. The method therefore results in the following system of nine coupled equations with $j=x, y, z$:
\begin{subequations}
\label{eq:Enskog_eqns}
\begin{align}
    & \dfrac{d \langle q_j^2 \rangle}{d t} - \dfrac{2}{m}\langle q_j p_j \rangle = 0, \\
    & \dfrac{d \langle p_j^2 \rangle}{d t} + 2m\omega_j^2\langle q_j p_j \rangle = \mathcal{C}[\Delta p_j^2], \\
    & \dfrac{d \langle q_j p_j \rangle}{dt} - \dfrac{1}{m}\left\langle p_j^2 \right\rangle + m\omega_j^2 \langle q_j^2 \rangle = 0.  
\end{align}
\end{subequations}

In these equations, collisions are incorporated through the integral
\begin{align}
    \mathcal{C}[\Delta\chi] = \frac{1}{N} & \int d^3 q d^3 p \int d\Omega_{p'} \frac{d\sigma}{d\Omega_{p'}} \nonumber \\
    &\times \int \frac{d^3p_1}{m} \norm{\boldsymbol{p} - \boldsymbol{p}_1} \left[ f' f_1' - f f_1 \right] \Delta\chi, \label{eq:Enskog_collision_integral}
\end{align}
where $\Delta\chi \equiv \chi^{\prime} + \chi_1^{\prime} - \chi - \chi_1$ denotes the amount by which $\chi$ changes during a collision event.  These nonlinear, coupled equations are known as the Enskog equations of change.

Notice that, in the absence of collisions ($\mathcal{C}=0$), these equations decouple along the three axes $q_j$. In this case, the normal modes of the Enskog equations along each axis $j$ come in two varieties: a breathing mode of angular frequency $2 \omega_j$, in which $\langle q_j^2 \rangle$ and $\langle p_j^2 \rangle$ are out of phase; and a stationary mode of frequency $\omega_j=0$, corresponding to the equilibrium configuration.  These modes are naturally modified by the presence of collisions, notably by shifting and broadening their resonant response functions.  This shift and broadening will, of course, depend on the dipolar properties of the cross section.

\subsection{\label{sec:AnisotropicScattering} Collision Integrals}

The Enskog equations are only complete when all $\mathcal{C}[\Delta \chi]$ collision terms are evaluated.  For the observables $\chi = x^2, y^2$ and $z^2$, $\mathcal{C}[\Delta \chi]$ vanishes, since the collision occurs at a given location, ${\bf q}={\bf q}^{\prime}$, hence $\Delta\chi=0$. Additionally, the observables $\chi = x p_x, y p_y$ and $zp_z$  also vanish, as can be seen by evaluating the collision integral in the center of mass frame. The only observables that contribute non-trivial collision integrals are then $\chi = p_x^2, p_y^2$ and $p_{z}^2$. The corresponding collision integrals in the Enskog formalism are given by
\begin{widetext}
\vspace{-1em}
\begin{subequations}
\label{eq:EnskogColl}
\begin{align}
    \mathcal{C}[\Delta p_x^2] \approx& \left( \dfrac{8 N}{15 \pi} \right) \Big( \frac{a_{\text{eff}}^2 m\overline{\omega}^3}{k_B T_0} \Big) \left[ \langle p_y^2 \rangle + \langle p_z^2 \rangle - 2\langle p_x^2 \rangle \right] \nonumber \\
    &+ a_d \left(\frac{64N}{105\pi}\right) \Big( \frac{a m \overline{\omega}^3}{k_B T_0} \Big) \Big[ \Big( \langle p_x^2 \rangle - \langle p_y^2 \rangle \Big) \cos (2 \alpha ) - 5\langle p_x^2 \rangle + 2\langle p_y^2 \rangle + 3\langle p_z^2 \rangle \Big] \nonumber \\
    &+ a_d^2 \left(\frac{4N}{315\pi}\right) \Big( \frac{m\overline{\omega}^3}{k_B T_0} \Big)  \Big[ \Big( \langle p_z^2 \rangle - \langle p_x^2 \rangle \Big) \cos (4 \alpha ) \nonumber \\
    &\quad\quad\quad\quad\quad\quad\quad\quad\quad\quad - 4\Big( \langle p_y^2 \rangle - \langle p_x^2 \rangle \Big) \cos (2 \alpha ) + 61\langle p_x^2 \rangle - 28\langle p_y^2 \rangle - 33\langle p_z^2 \rangle \Big], \label{subeq:EnskogColl1} \\
    \mathcal{C}[\Delta p_y^2] \approx& \left( \dfrac{8 N}{15 \pi} \right) \Big( \frac{a_{\text{eff}}^2 m \overline{\omega}^3}{k_B T_0} \Big) \left[ \langle p_x^2 \rangle + \langle p_z^2 \rangle - 2\langle p_y^2 \rangle \right] \nonumber \\
    &-  a_d\left(\frac{64N}{105\pi}\right) \Big( \frac{a m \overline{\omega}^3}{k_B T_0} \Big) \Big[ \Big( \langle p_x^2 \rangle - \langle p_z^2 \rangle \Big) \cos (2 \alpha ) - 2\langle p_x^2 \rangle + 4\langle p_y^2 \rangle - 2\langle p_z^2 \rangle \Big] \nonumber \\
    &+ a_d^2\left(\frac{16N}{315\pi}\right) \Big( \frac{m\overline{\omega}^3}{k_B T_0} \Big)  \Big[ \Big( \langle p_x^2 \rangle - \langle p_z^2 \rangle \Big) \cos (2 \alpha ) - 7\langle p_x^2 \rangle + 14\langle p_y^2 \rangle - 7\langle p_z^2 \rangle \Big], \label{subeq:EnskogColl2} \\
    \mathcal{C}[\Delta p_z^2] \approx& \left( \dfrac{8 N}{15 \pi} \right) \Big( \frac{a_{\text{eff}}^2 m \overline{\omega}^3}{k_B T_0} \Big) \left[ \langle p_x^2 \rangle + \langle p_y^2 \rangle - 2\langle p_z^2 \rangle \right] \nonumber \\
    &+ a_d\left(\frac{64N}{105\pi}\right) \Big( \frac{a m \overline{\omega}^3}{k_B T_0} \Big) \Big[ \Big( \langle p_y^2 \rangle - \langle p_z^2 \rangle \Big) \cos (2 \alpha ) + 3 \langle p_x^2 \rangle + 2\langle p_y^2 \rangle - 5 \langle p_z^2 \rangle \Big] \nonumber \\
    &+ a_d^2\left(\frac{4N}{315\pi}\right) \Big( \frac{m\overline{\omega}^3}{k_B T_0} \Big) \Big[ \Big( \langle p_x^2 \rangle - \langle p_z^2 \rangle \Big) \cos (4 \alpha ) \nonumber \\
    &\quad\quad\quad\quad\quad\quad\quad\quad\quad\quad - 4 \Big( \langle p_y^2 \rangle - \langle p_z^2 \rangle \Big) \cos (2 \alpha ) - 33\langle p_x^2 \rangle - 28\langle p_y^2 \rangle + 61\langle p_z^2 \rangle \Big], \label{subeq:EnskogColl3}
\end{align}
\end{subequations}
\end{widetext}
to linear order in $\langle p_j^2 \rangle$. 

This result is given in terms of an effective length scale that combines the scattering length and the dipole length, via $a_{\text{eff}}^2 \equiv 2\left( a^2 - {4a a_d / 3} + {4a_d^2 / 9} \right)$, and in terms of the geometric mean of trap frequencies, $\overline{\omega}^3 = \omega_{\perp}^2 \omega_{z,0}$. Details of this derivation can be found in Appendix~\ref{app:DipolarCollInt}. 

The collision integrals (\ref{eq:EnskogColl}) are  decomposed into terms of increasing orders in $a_d$, emphasizing the anisotropic collisional effects.  Note that these results match those of the isotropic scatterers in \cite{Guerey99_PRA} when we set $a_d = 0$ and return to isotropic scattering.

\subsection{\label{sec:NumericalSim} Numerical Simulations}

The Enskog equations have the advantage of being simple to implement and, in principle, to provide analytical insight. However, they are restricted to the limit of weak drive and assume that the phase space distribution remains nearly Gaussian. It is therefore useful to establish a more robust numerical method that is not limited to the perturbative regime.  

Numerical time evolution of the gas is performed by first approximating the phase-space distribution function with a discrete ensemble of particles with phase space locations $(\boldsymbol{q}_k,\boldsymbol{p}_k)$, randomly sampled from the initial distribution $f(\boldsymbol{q}, \boldsymbol{p}, 0)$.  This gives the distribution
\begin{equation}
    f(\boldsymbol{q}, \boldsymbol{p}) \approx \xi \sum_{k = 1}^{N_T} \delta^3(\boldsymbol{q} - \boldsymbol{q}_k) \delta^3(\boldsymbol{p} - \boldsymbol{p}_k),
\end{equation}
where $\xi = N / N_t$ and $N_t$ is the number of numerically simulated particles which we refer to as ``test particles". All simulations performed for this work take $N_t = N$, which has been proven to provide good stochastic convergence. This allows us to drop further use of the variables $N_t$ and $\xi$.

Trajectories of these test particles are computed by following their Hamiltonian dynamics with the equations of motion
\begin{subequations}
\label{eq:N2L}
\begin{align}
    & \dot{\boldsymbol{q}} = \dfrac{\boldsymbol{p}}{m} \\
    & \dot{\boldsymbol{p}} = -\grad_q U\left( \boldsymbol{q}; t \right),
\end{align} 
\end{subequations}
and are solved numerically using a fourth-order Runge-Kutta method (RK4) \cite{Ixaru}. We chose the RK4 over a symplectic integrator (e.g. velocity Verlet) due to the explicit time-dependence in the Hamiltonian. This leads to changes in the phase-space volume which the RK4 makes no assumptions about. Moreover, energy drifts typically associated to the RK4 are negligible for the time intervals we simulate in this work. The state-vector under numerical integration is defined to be
\begin{align}
    \boldsymbol{y} = \begin{pmatrix}
        {\boldsymbol{q}} \\
        {\boldsymbol{p}}
    \end{pmatrix},
\end{align}
with the numerical time-step $\Delta t$ for this integration scheme chosen to be much smaller than the mean time interval between collision.

Collisions are included using the direct simulation Monte Carlo (DSMC) method \cite{Bird1970, DSMC}. For the present situation of a cold, dipolar gas, this method has been implemented previously  \cite{Sykes15_PRA}, and we follow this implementation here.  We construct a discrete spatial grid of cubic grid cells with constant volume determined by $V_{\text{cell}} = {\beta / n_{\text{ave}}}$. $\beta$ is an initialized guess of the number of particles that would be contained in each cell and $n_{\text{ave}}$ is the average number density
\begin{equation}
    n_{\text{ave}} = \frac{N}{ (2\pi)^{3/2} \sigma_x \sigma_y \sigma_z },
\end{equation}
where $\sigma_j$ is the standard deviation of particle positions along axis $j$. $\beta$ is a free-parameter of the simulation that can be optimized. The number of grid cells varies with each time-step depending on the position of the particles. 

The simulated particles are binned into each grid cell based on their positions, for which collision processes are performed in the following 2 main steps.

\paragraph{Determination of Collisions}

A collision occurs when the probability  
\begin{align}
    P_{ij}(\text{collision}) = \left( \frac{\Delta t}{m \Delta V} \right) |\boldsymbol{p}_{\text{rel}}| \: \sigma (\hat{\boldsymbol{p}}) 
\end{align}
exceeds a random number $R$ sampled from a uniform distribution $\mathcal{U}(0, 1)$, where $\boldsymbol{p}_{\text{rel}} = \boldsymbol{p}_i - \boldsymbol{p}_j$ and $\sigma(\hat{\boldsymbol{p}})$ (total scattering cross section) has the closed-form \cite{Bohn14_PRA}
\begin{align}
    \sigma(\boldsymbol{p}_{\text{rel}}) = & \frac{\pi}{9} \big[ 72a^2 - 24 a_d a \left(1 - 3\cos^2\eta\right) \nonumber \\
    & + \left( 11 - 30\cos^2\eta + 27\cos^4\eta \right) a_d^2 \big]  \label{total_cross_sections_bosons}
\end{align}
where $\eta = \cos^{-1}(\hat{\boldsymbol{p}} \cdot \hat{\varepsilon})$ is the angle between the relative momentum and the dipole-alignment axis. 

\paragraph{Updating Dynamical Variables}

If a collision is said to proceed (from step 1), the post-collision momenta of the particles are then computed using the rejection sampling algorithm described in \cite{Sykes15_PRA}.

More efficient numerical schemes exist in the literature such as those which employ locally adaptive methods \cite{Wade_PRA}. However, the size of experiments we simulate permits our current implementation to achieve excellent numerical convergences in reasonable computational time. We leave the task of optimization for subsequent iterations of our simulation program, which for instance, could be used to investigate far-from-equilibrium systems. 

Once the phase space locations $(\boldsymbol{q}_k,\boldsymbol{p}_k)$ are given for each test particle, the averages required to compute observables are found in a straightforward way:
\begin{subequations}
\begin{align}
    & \langle q_j^2 \rangle = \frac{1}{N} \sum_{k=1}^{N} q_j^{(k)}(t)^2, \\
    & \langle p_j^2 \rangle = \frac{1}{N} \sum_{k=1}^{N} p_j^{(k)}(t)^2,
\end{align}
\end{subequations}
where the sum over $k$ runs over all simulated particles. 

\section{\label{sec:results} Results}

The chemical element with the largest magnetic dipole moment is dysprosium, so it is the natural candidate for our study. In particular, all atomic species parameters adopted in this work are those of $^{164}$Dy unless otherwise specified. Furthermore, we choose trapping frequencies $\omega$ and number of particles $N$ such that the average number density of the gas lands us in the weakly hydrodynamic regime. This is where the mean collision rate is larger but on the order of the trapping frequency, $n_{\text{ave}} \sigma \langle v \rangle \gtrsim \omega / 2\pi$. In this regime the influence of collision will definitely be seen, while the influence of anisotropic scattering is unlikely to be washed out by multiple collisions.

To subject the gas to a driving that is perturbative, we set the relative modulation amplitude in Eq.~(\ref{eq:time_varying_trap}) equal to $\delta = 0.05$. The parameter values used to produce the results of the simulation are given in Table~\ref{tab:system_parameter}. In particular, for most simulations we incorporate the actual $s$-wave scattering length of $^{164}$Dy, as determined in Ref.~\cite{Tang15_PRA}.

\vspace{-1em}
\begin{table}[H]
\caption{\label{tab:system_parameter} 
Table of parameter values with which the system is initialized. Da $= 1.661 \times 10^{-27}$ kg stands for Dalton (atomic mass unit), $a_{0} = 5.292 \times 10^{-11}$ m is the Bohr radius and $\mu_B = 9.274 \times 10^{-24}$ J/T is the Bohr magneton. }
\begin{ruledtabular}
\begin{tabular}{l c c c}
\multicolumn{1}{c}{\textrm{Parameter}} & \multicolumn{1}{c}{\textrm{Symbol}} & \multicolumn{1}{c}{\textrm{Value}} & \multicolumn{1}{c}{\textrm{Unit}} \\
    \colrule
    Number of particles, & $N$ & 80, 000 & -- \\
    Atomic mass number, & $A$ & 164 & Da \\
    Magnetic moment & $\mu$ & 10 & [$\mu_B$] \\
    s-wave scattering length, & $a$ & 92 & [$a_{0}$] \\
    Dipole length, & $a_d$ & 199 & [$a_{0}$] \\
    Initial gas temperature, & $T_0$ & 426 & nK \\
    Axial trapping frequency, & $\omega_{z, 0}$ & $2\pi\times 40$ & Hz \\
    Radial trapping frequency, & $\omega_{\perp}$ & $2\pi\times 400$ & Hz
\end{tabular}
\end{ruledtabular}
\end{table}

\subsection{\label{sec:EnskogVsDSMC} Enskog and DSMC Results for Anisotropic Pseudotemperature}

\begin{figure}[h]
    \centering
    \includegraphics[width=0.965\textwidth]{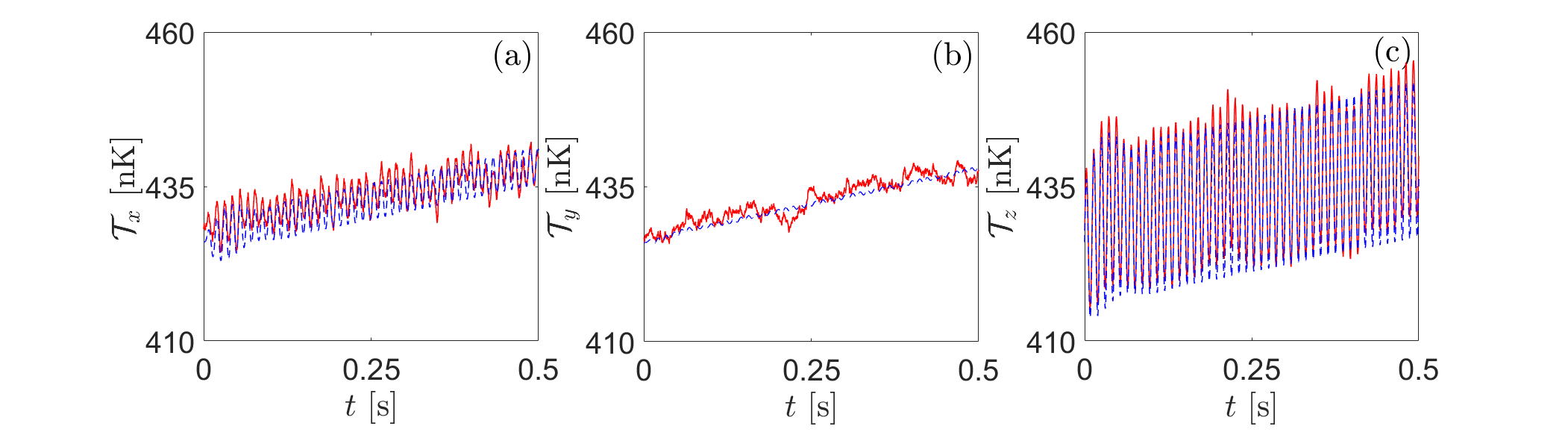}
    \caption{Pseudotemperatures of a dipolar gas periodically driven along $z$: comparison between solutions from a numerical DSMC simulation and the Enskog equations for \textbf{off-resonant} driving at $\Omega = 2\pi \times 90$ [Hz]. (a) $\mathcal{T}_x$; (b) $\mathcal{T}_y$; and (c) $\mathcal{T}_z$. The numerical solution is shown by the solid (red) line; the Enskog solutions are shown by the dashed (blue) line. The Enskog solutions are seen to model the heating rate, oscillation phase and amplitude accurately when compared to the numerical results. These results were obtained with $\alpha = 90^{\circ}$. }
    \label{fig:a92_ad199_Omega90}
\end{figure}

\begin{figure}[h]
    \centering
    \includegraphics[width=0.965\textwidth]{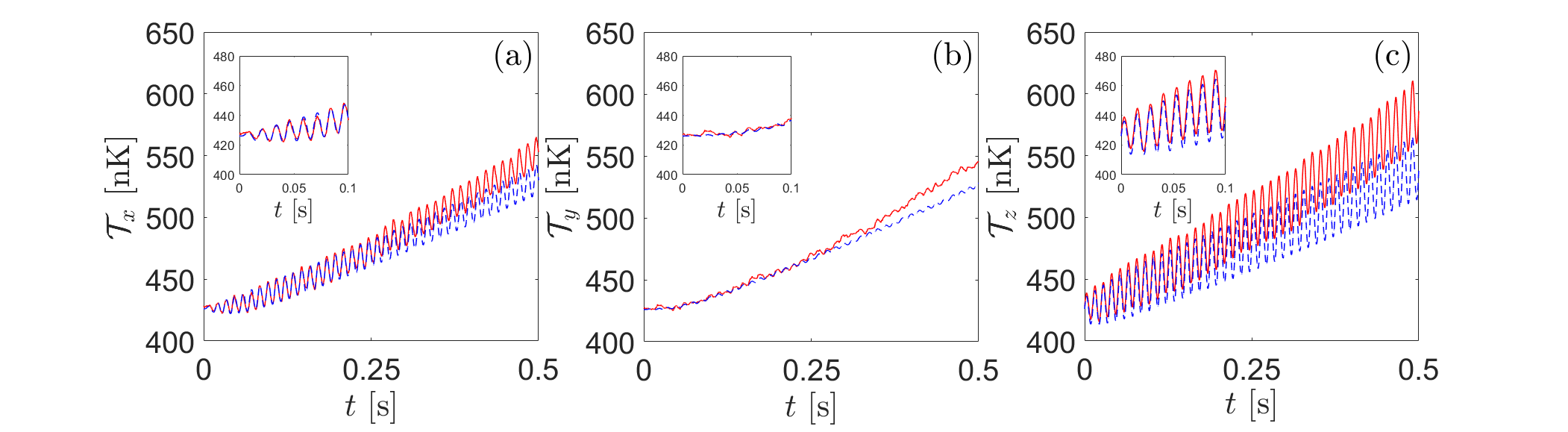}
    \caption{Pseudotemperatures of a dipolar gas periodically driven along $z$: comparison between solutions from a numerical DSMC simulation and the Enskog equations for \textbf{close-to-resonant} driving at $\Omega = 2\pi \times 90$ [Hz]. (a) $\mathcal{T}_x$; (b) $\mathcal{T}_y$; and (c) $\mathcal{T}_z$. The numerical solution is shown by the solid (red) line; the Enskog solutions are shown by the dashed (blue) line. The Enskog solutions are seen to closely follow the numerical results within the linear regime ($t$ up to $0.1$s shown in the embedded plots) but deviate once the response becomes nonlinear. These results were obtained with $\alpha = 90^{\circ}$. }
    \label{fig:a92_ad199_Omega80}
\end{figure}

As the drive is along the $z$ axis with trap frequency $\omega_z = 2\pi \times 40$ Hz, it is expected that the gas will resonate at a frequency $\sim 2 \omega_z \sim 2\pi \times 80$ Hz.  The basic time-dependent response of the gas to weak periodic perturbation is given in Figs.~\ref{fig:a92_ad199_Omega90} and \ref{fig:a92_ad199_Omega80} for non-resonant and resonant dives respectively. In both illustrative cases, the dipoles are polarized in the $x$ direction, with $\alpha = 90^{\circ}$.

Fig.~\ref{fig:a92_ad199_Omega90} depicts off-resonant driving, at the driving frequency $\Omega = 2\pi \times 90$ Hz. The three panels show the pseudotemperatures for the three Cartesian axes.  The blue dashed curves give the results of numerically solving the Enskog equations, while the red curve is the result from the DSMC. In all three panels the agreement between the methods is quite good, validating the Enskog equations as derived. In all cases the DSMC produces noisier results from persistent collisions, as expected.

The pseudotemperture in all three directions rises at the same mean rate, $\sim 27$ [nK/s], but in qualitatively different ways. The greatest oscillation about the mean rise is, naturally, in the $z$ axis along which the drive is applied.  Fluctuations about the mean rise in the other directions are the result of collisions. Atoms that are driven primarily along the $z$ axis, upon scattering, are more likely to scatter into the $x$ direction (along the polarization axis) than in the $y$ direction, perpendicular to the polarization axis, according to the differential cross section. The anisotropic nature of the dipoles and their scattering therefore plays a role in the initial rate of energy distribution in the gas.

\onecolumngrid

\twocolumngrid

In Figure \ref{fig:a92_ad199_Omega80} the gas is driven much closer to resonance, at $\Omega = 2 \pi \times 80$ Hz, hence the heating rate is much faster.  Here the results of the Enskog approximation and the full dynamics in the DSMC are in agreement, but only until $\sim 0.25$ seconds, where the response of the gas becomes nonlinear. This circumstance sets a limit over which the linear response theory is expected to apply to this system.  The general trend still holds however, that deviations from the mean heating rate are greatest in the $z$ direction and least in the $y$ direction.  Also of note is the transient delay in the non-driven directions: while the $z$ direction begins heating immediately, the $x$ and $y$ pseudotemperatures require several ($\sim 5$) trap oscillations ($\sim 16$ mean collision times)  before beginning to heat at a comparable to the $z$ axis.

\subsection{\label{sec:NormalModes} Normal Modes of a Dipolar Gas}

The numerical results of the previous section verify that there is a limit with sufficiently weak driving and sufficiently short times where the response of the gas is linear and is adequately described by the Enskog version of the theory.  Hereafter, we exploit this version to explore the normal modes of the gas. To this end, we seek eigensolutions of the Enskog equations in the absence of driving, setting $\delta=0$ in (\ref{eq:time_varying_trap}).

The linearized Enskog equations of change in Eqs.~(\ref{eq:Enskog_eqns}) and (\ref{eq:EnskogColl}) constitute a linear system of equations for nine dynamical quantities that can be written in terms of a nine-dimensional state-vector
\begin{align}
    \boldsymbol{\xi}(t) = \Big[
        & m^2\omega_{z, 0}^2\langle z^2 \rangle, \:\:
        \langle p_z^2 \rangle, \:\:
        m \omega_{z, 0}\langle z p_z \rangle, \nonumber \\
        & \quad\quad m^2 \omega_{\perp}^2\langle y^2 \rangle, \:\:
        \langle p_{y}^2 \rangle, \:\:
        m \omega_{\perp} \langle y p_y \rangle, \nonumber \\
        & \quad\quad\quad\quad m^2 \omega_{\perp}^2 \langle x^2 \rangle, \:\:
        \langle p_{x}^2 \rangle, \:\:
        m \omega_{\perp} \langle x p_x \rangle \Big]^T. \label{eq:linear_Enskog_state_vector}
\end{align}

Cast in terms of this vector, the Enskog equations can be written in the succinct form 
\begin{align}
    \dot{\boldsymbol{\xi}}(t) = \boldsymbol{\Phi}_0 \boldsymbol{\xi}(t), \label{eq:Enskog_linsys}
\end{align}
where $\boldsymbol{\Phi}_0$ is a matrix of coefficients with units of frequency, which can be read off from  Eqs.~(\ref{eq:Enskog_eqns},\ref{eq:EnskogColl}). This allows the identification of intrinsic normal modes (often referred to as collective oscillations \cite{Guerey99_PRA, Vichi99_PRA, Vichi00_JLTP, PedriStringari}) of this system without time-dependent driving. The normal mode solutions are found by means of the ansatz
\begin{align}
    \boldsymbol{\xi}(t) = \boldsymbol{\xi}_0 + \boldsymbol{\xi}_{\omega} e^{i\omega t},
    \label{eq:matrix_Enskog}
\end{align}
where $\boldsymbol{\xi}_0$ is the equilibrium solution, $\boldsymbol{\xi}_{\omega}$ is a vector of relative amplitudes and $\omega = \omega_r + i\Gamma$ is the complex-valued frequency, with real part $\omega_r$ being the frequency of oscillation and imaginary part $\Gamma$ the damping rate. 
Substituting into (\ref{eq:matrix_Enskog}), the normal mode frequencies satisfy the eigenvector equation 
\begin{align}
    \boldsymbol{\Phi}_0\boldsymbol{\xi}_{\omega} = i\omega \boldsymbol{\xi}_{\omega}
\end{align}

Solving the eigensystem, we find two varieties of unique and dynamical eigenmodes: 1) three oscillatory modes with oscillation frequencies close to the trap frequencies, that are damped due to cross-dimensional rethermalization; and 2) two overdamped modes with $\omega_r=0$, that relax to equilibrium without oscillating. We refer to the latter variety as ``melting modes''. These solutions are generalizations of the zero-frequency modes  discussed at the end of Sec.~\ref{sec:MethodOfAverages}, with the addition of coupling between the axes through collisions. 

These solutions are presented with plots depicting the eigenfrequency solutions $\omega_r$ and $\Gamma$, as a function of the dipolar tilt angle $\alpha$, along with the associated time-evolution of the spatial eigenvector components $\Re\{\langle q_j^2 \rangle_{\omega} e^{i\omega t}\}$, at select values of $\alpha$. In all cases, the eigenvectors are normalized to unity at time $t=0$.

\begin{figure}[h]
    \centering
    \includegraphics[width=\columnwidth]{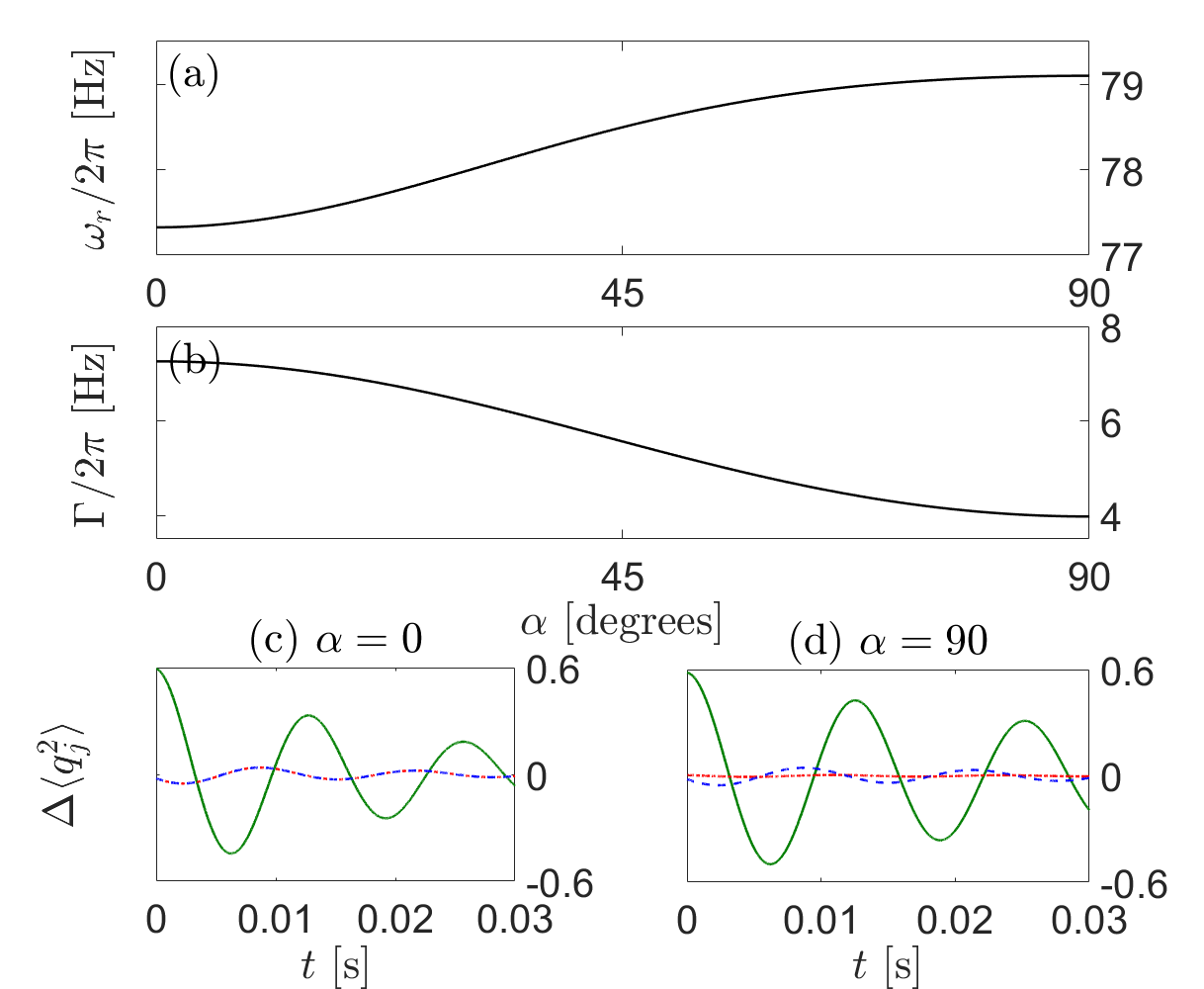}
    \caption{The normal mode along $z$. Panels (a) and (b): real $\omega_r$, and imaginary $\Gamma$, parts of the eigenfrequency respectively as $\alpha$ is varied from $0$ to $90^{\circ}$. Panels (c) and (d): time-evolution of the associated normal modes for $\alpha = 0^{\circ}, 90^{\circ}$, where the solid (green) curve denotes the relative amplitude of $\langle z^2 \rangle$, the dash-dotted (red) curve $\langle y^2 \rangle$ and the dashed (blue) curve $\langle x^2 \rangle$. The axial oscillations of this solution initially dominate over the radial oscillations. }
    \label{fig:Mode3}
\end{figure}

Fig.~\ref{fig:Mode3} illustrates the mode that oscillates primarily in the $z$ direction, hence has frequency $\sim 2 \omega_z \sim 2 \pi \times 80$ Hz (Fig.~\ref{fig:Mode3}\textcolor{blue}{a}). The frequency of this mode rises only slightly as the dipole is tilted from along the $z$ axis ($\alpha=0$ to perpendicular to this axis $\alpha=90^{\circ}$).  By contrast, tilting the dipole has a dramatic effect on the damping rate, cutting it nearly in half as $\alpha$ is tuned from $0$ to $90^{\circ}$ (Fig.~\ref{fig:Mode3}\textcolor{blue}{b}). This is a consequence of the changing differential cross section as $\alpha$ is varied.

The character of the mode is illustrated by the time traces of relative amplitudes about thermal equilibrium (denoted $\Delta \langle q_j^2 \rangle$) in Figures \ref{fig:Mode3}\textcolor{blue}{c,d}. The principal motion defining this mode is excitation in the $z$ direction. Hence the modest excitations in $x$ and $y$ are driven by collisions. When $\alpha = 0$ (Fig.~\ref{fig:Mode3}\textcolor{blue}{c}), cylindrical symmetry holds, and the differential cross section sends atoms equally into the $x$ and $y$ excitations.   By contrast, when $\alpha = 90^{\circ}$ the differential cross section favors scattering into the $x$ direction, and scatters hardly anything into the $y$ direction.  Because of this, the overall scattering rate is reduced, and therefore so is the damping rate.

\begin{figure}[h]
    \centering
    \includegraphics[width=\columnwidth]{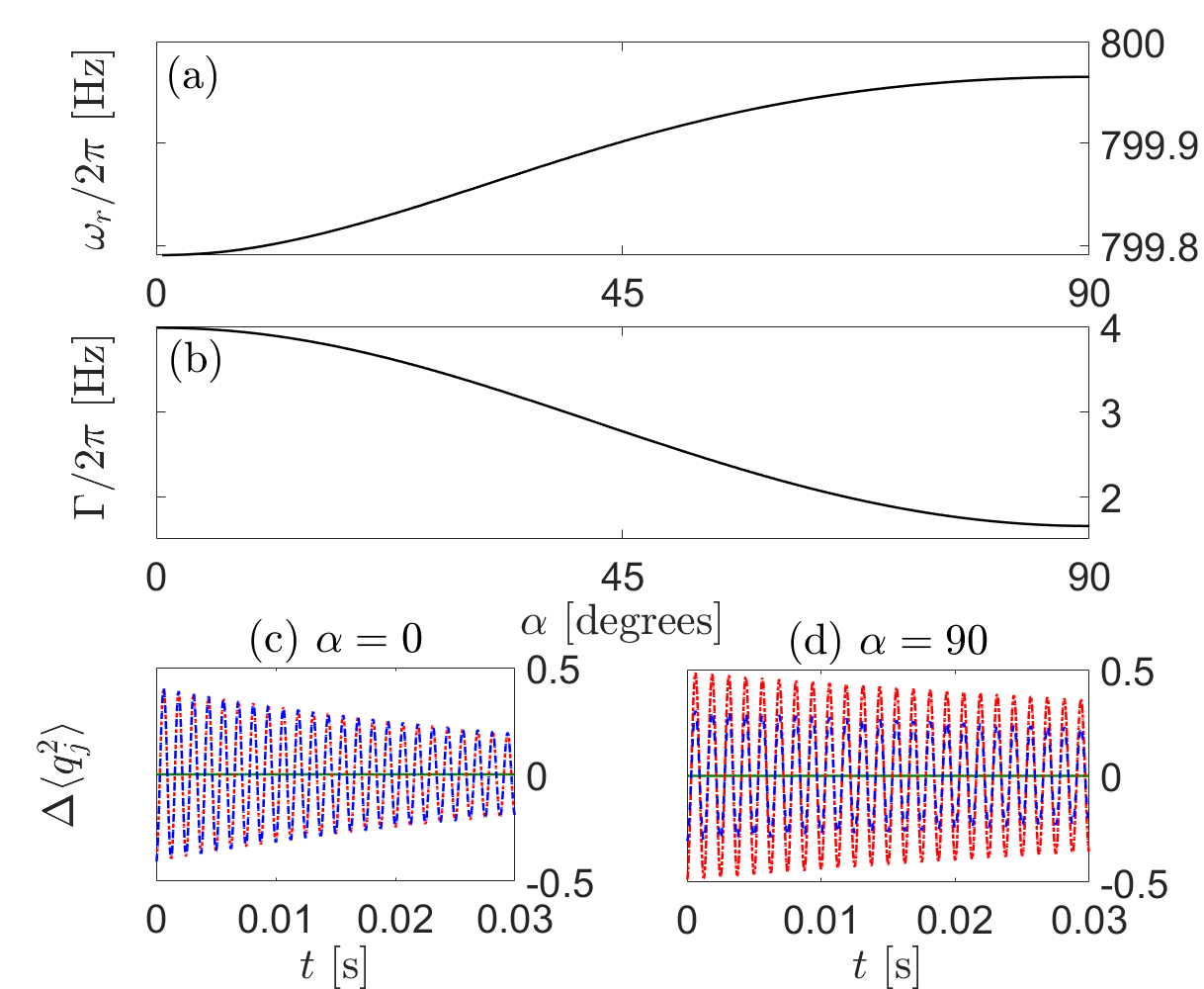}
    \caption{Approximate breathing mode in $x$-$y$. Data format, colors and markers follow that in Fig.~\ref{fig:Mode3}. The radial oscillations of this solution initially dominate over the axial oscillations and are in-phase, creating an approximate radial breathing mode. }
    \label{fig:Mode1}
\end{figure}

\begin{figure}[h]
    \centering
    \includegraphics[width=\columnwidth]{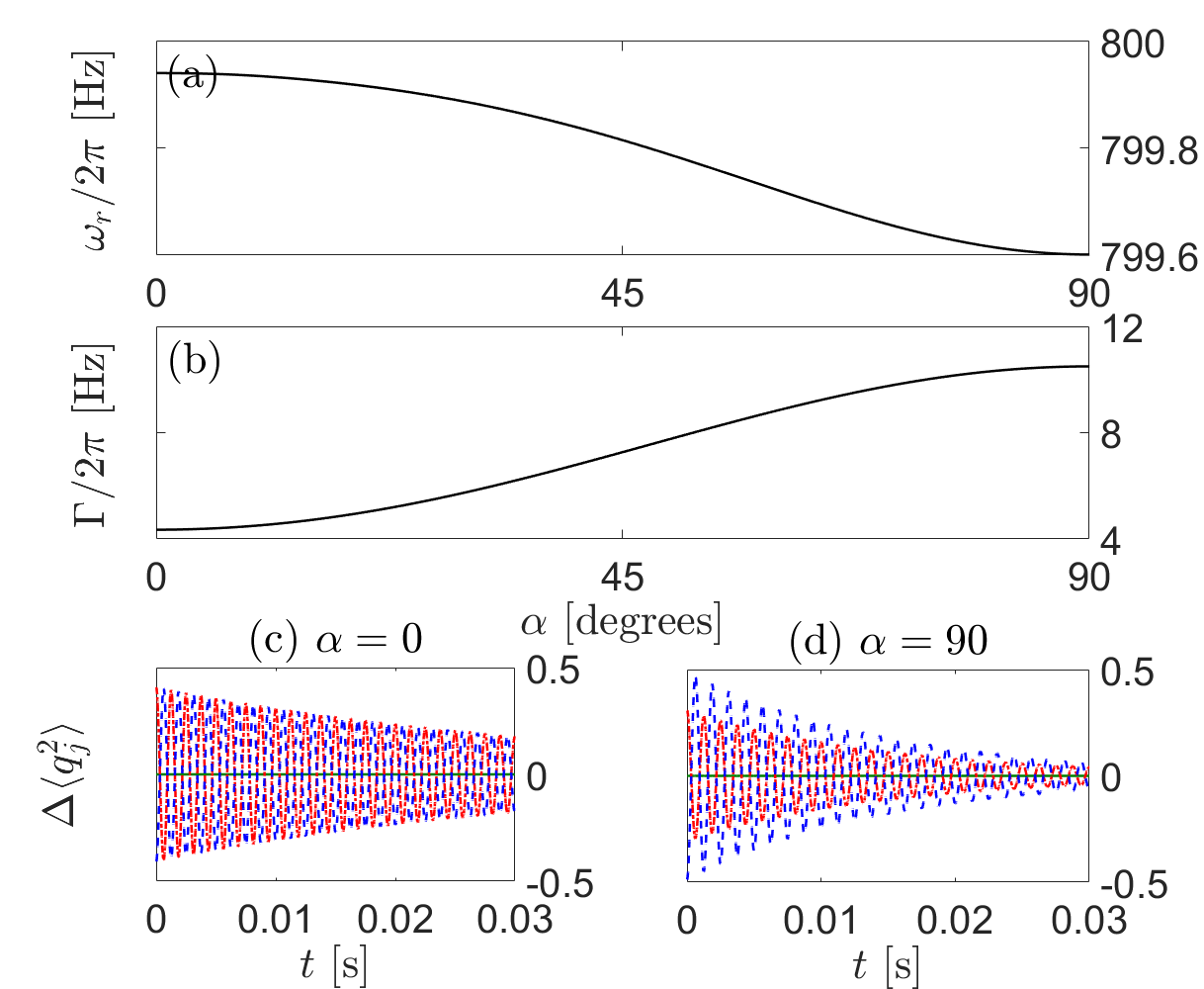}
    \caption{Approximate quadrupole mode in $x$-$y$. Data format, colors and markers follow that in Fig.~\ref{fig:Mode3}. The radial oscillations of this solution initially dominate over the axial oscillations and are $\pi$ radians out-of-phase, creating an approximate radial quadrupole mode. }
    \label{fig:Mode2}
\end{figure}

The other two oscillatory modes have state-vector amplitudes which are initially dominant in the radial directions, shown in Figs.~\ref{fig:Mode1} and \ref{fig:Mode2}. These modes accordingly have resonant frequencies $\omega_r \sim 2\omega_{\perp} \sim 2 \pi \times 800$ Hz. The mode in Fig.~\ref{fig:Mode1} has radial oscillations in-phase, much like a breathing mode in the $x$-$y$ plane.  When $\alpha=0$ the amplitudes in the two directions are equal (Fig.~\ref{fig:Mode1}\textcolor{blue}{c}). However, the breathing is distorted when the dipole alignment axis is tilted off the trap axis of symmetry, resulting (Fig.~\ref{fig:Mode1}\textcolor{blue}{d}). Along with this, the damping rate decreases as the dipoles are tilted from $\alpha = 0$ to $\alpha = 90^{\circ}$. This differentiation in amplitude is also seen in Fig.~\ref{fig:Mode2}, whereby in this solution, the radial oscillations are $\pi$ radians out-of-phase, likened to a radial quadrupole mode. In contrast to the breathing modes, the damping rate in this case increases as $\alpha$ grows from $0$ to $90^{\circ}$.

\begin{figure}[h]
    \centering
    \includegraphics[width=\columnwidth]{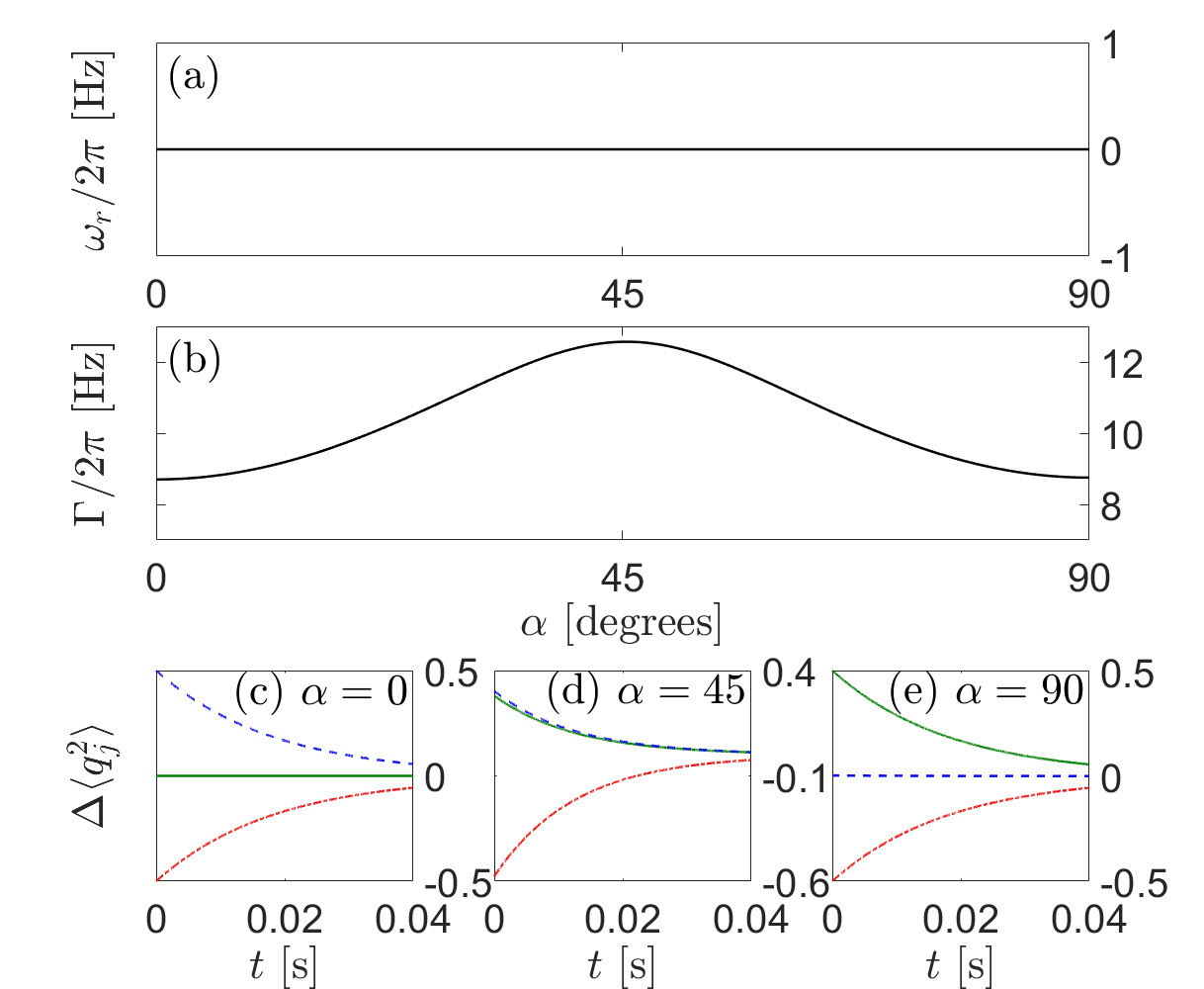}
    \caption{First melting mode solution. Data format, colors and markers follow that in Fig.~\ref{fig:Mode3}. The $\langle x^2 \rangle$ and $\langle z^2 \rangle$ amplitudes appear to cross and exchange positions as $\alpha$ goes from $0$ to $90^{\circ}$, breaking the radial symmetry. This mode solution has no oscillatory component ($\omega_r = 0$), resulting in purely damping dynamics. }
    \label{fig:Mode4}
\end{figure}

\begin{figure}[h]
    \centering
    \includegraphics[width=\columnwidth]{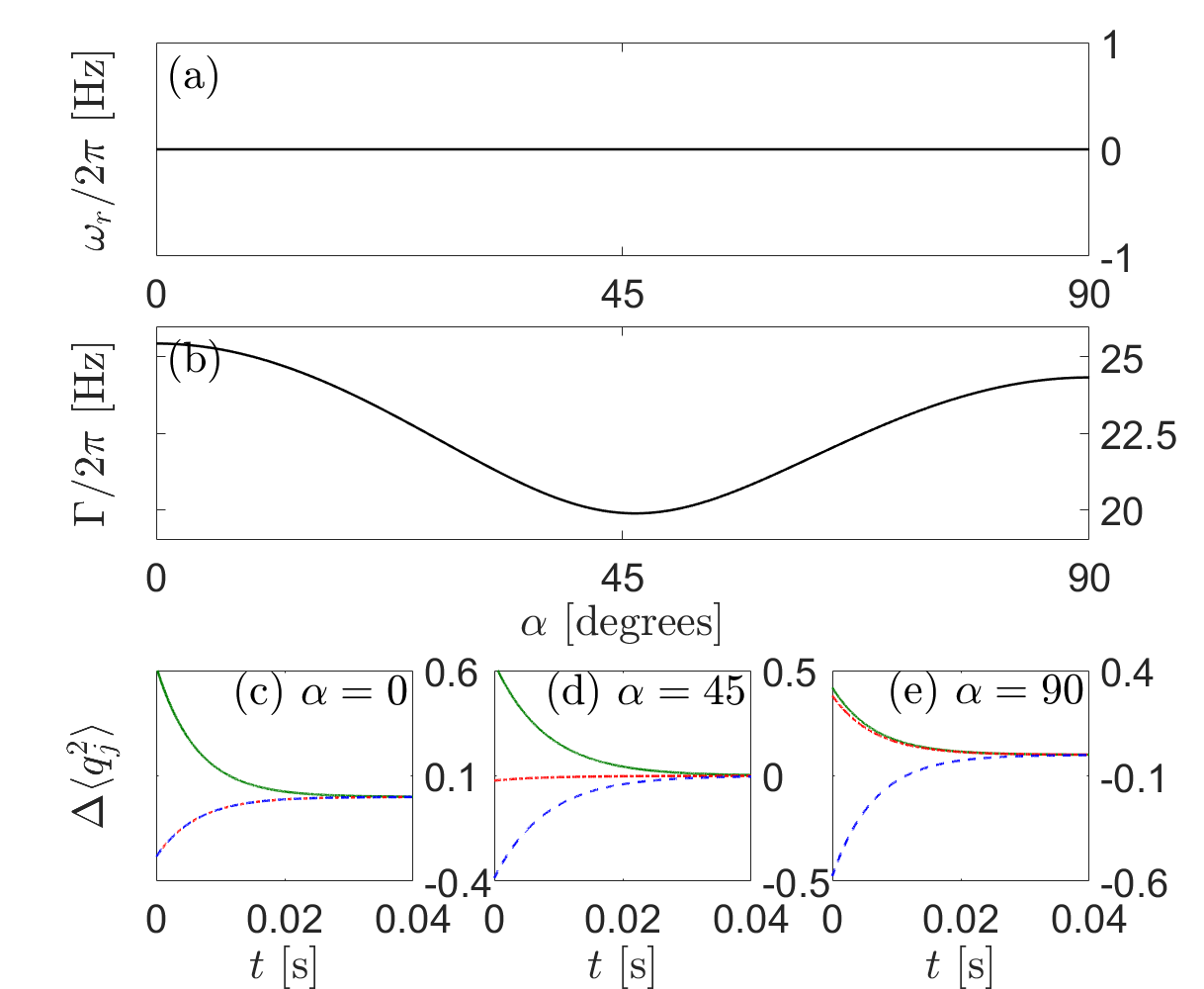}
    \caption{Second melting mode solution. Data format, colors and markers follow that in Fig.~\ref{fig:Mode3}.This solution also has $\omega_r = 0$, where symmetry between the radial axes is again broken when $\alpha$ increases from $0$ to $90^{\circ}$.  }
    \label{fig:Mode5}
\end{figure}

The second variety of normal mode solutions are those with no oscillations ($\omega_r = 0$), which mean that these modes when excited, strictly relax to thermal equilibrium with no additional dynamics. These are presented in Figs.~\ref{fig:Mode4} and \ref{fig:Mode5}. The damping rates $\Gamma$, of these solutions have a local extremum at $\alpha \approx 45^{\circ}$, so plots of their time-evolution are given for $\alpha = 45^{\circ}$, in addition to $\alpha = 0, 90^{\circ}$.

The dynamic action of the melting modes is highly dependent on the initial condition.  Consider the first mode with tilt angle $\alpha=0$, depicted in  Fig.~\ref{fig:Mode4}\textcolor{blue}{c}. The initial condition requires the gas to be slightly compressed in the $y$ direction (red), and extended in the $x$ direction (blue), much like a radial quadrupole mode.  However, for the particular distortion shown, these initial amplitudes simply decay back to the equilibrium size of the cloud, with no oscillation at all.  Interestingly, as the dipoles are tilted and $\alpha$ proceeds through $45^{\circ}$ to $90^{\circ}$, the initial distortion evolves into a slight compression of the gas in the $y$ direction (red), coupled with a slight extension in the $z$ direction (green).

The other melting mode is depicted in Figure \ref{fig:Mode5}.  In this somewhat more complicated mode, the initial condition for $\alpha = 0$ requires a slight expansion in $z$ and a slight contraction in both $x$ and $y$ (Fig.~\ref{fig:Mode5}\textcolor{blue}{c}). As the dipole is tilted, the required distortion in the $y$ direction (red), changes from an initial compression to an initial extension.

These modes could in principle be realized experimentally by means of a trap quench, similar to that done in \cite{Sykes15_PRA}, that sets the appropriate initial shape of the gas. We suspect however, that this would be rather difficult to accomplish for several reasons. Firstly, DSMC simulations have made apparent a high sensitivity to initial conditions, in that these modes can only be excited through an initial configuration very close to thermal equilibrium. This raises concerns on signal to noise ratios when these collective excitations are to be measured. Additionally, fluctuations in the positions and momentum of the atoms result in the excitation of other modes, causing inevitable oscillations and other transient dynamics en route to thermalization. These issues present an intriguing problem, to be investigated in future works.

\section{\label{sec:LinearResponse} Linear Response of a Dipolar Gas }

In the previous section we considered anisotropy from the perspective that motion can be different in the three Cartesian directions, both in the normal modes and in the response to a weak periodic drive.  Here instead, we show how the collective heating of the driven gas is a function of the dipole orientation.  We also note that this anisotropy is strongly affected by the value of the $s$-wave scattering length.

The customary response function for resonantly driven systems is the transmissibility, defined as the ratio of the output response to input driving amplitudes. Having linearized the Enskog equations allows us to compute the transmissibility from a linear frequency response function. The periodic trap modulations are modeled by a change in the trap frequency according to Eq.~(\ref{eq:time_varying_trap}), resulting in additional terms in the Enskog Eqs.~(\ref{eq:Enskog_eqns}).  The modified equations can be written as
\begin{align}
\begin{split}
    & \dot{\boldsymbol{\xi}}(t) = \Phi_0 \boldsymbol{\xi}(t) + \widetilde{\Phi}(t)\boldsymbol{\xi}(t),
\end{split}
\end{align}
where $\Phi_0$ is defined above as in the derivation of normal modes, and
\begin{align}
    \widetilde{\Phi}(t)\boldsymbol{\xi}(t) =  \big[ & 0, - 2m \omega_{z,0}^2  \langle z p_z \rangle , - m^2\omega_{z,0}^2 \langle z^2 \rangle, \nonumber\\
     & 0, 0, 0, 0, 0, 0 \big]^T \delta \sin(\Omega t).
\end{align}

Considering just the linear response by taking the limit of $t \rightarrow 0$, we approximate the time-dependent vector using values of the Enskog state variables at thermal equilibrium, $\langle zp_z\rangle_{0} = 0, \langle z^2\rangle_{0} = k_B T_0 / (m\omega_{z, 0}^2)$. This permits the system to be treated as driven by state-variable independent inputs, satisfying
\begin{align}
    \left[ \frac{d}{dt} - \boldsymbol{\Phi}_0 \right]
    \boldsymbol{\xi}(t) =&  -  m k_B T_0 \delta \sin(\Omega t) \boldsymbol{u}_3 \\
    \equiv & -h \sin(\Omega t) \boldsymbol{u}_3,
\end{align} 
where $\boldsymbol{u}_3 $ is a vector with $1$ in the third entry, and zeros in the remaining eight entries. This formulation also identifies the strength of the drive as $h = m k_B T_0 \delta$.  Our goal is to find the response relative to this drive amplitude. This now allows us to derive a linear response function (a.k.a. Greens function) in the time-domain. To do so, we consider the impulse response
\begin{align}
    \left[ \frac{d}{dt} - \boldsymbol{\Phi}_0 \right]\boldsymbol{G}(t- t') = \mathbf{I} \delta(t - t'),
\end{align}
with $\boldsymbol{G}(t- t')$ being the response matrix, $\mathbf{I}$ the identity matrix and $\delta(t)$ the Dirac-delta function.  
Utilizing the method of Laplace transforms,  the solution to this equation takes the form
\begin{align}
    \boldsymbol{G}(t- t') = \Theta(t - t') \exp\left[ \boldsymbol{\Phi}_0 (t - t') \right],
\end{align}
where $\Theta(t)$ is the Heaviside step function. The response matrix in frequency space is then obtained by taking a Fourier transform
\begin{align}
    \tilde{\boldsymbol{G}}(\Omega) = \mathcal{F}\left\{ \boldsymbol{G}(t) \right\} = \left( i\Omega - \boldsymbol{\Phi}_0 \right)^{-1},
\end{align}
whose real part constitutes a reactance matrix and whose imaginary part constitutes an dissipation matrix. 

At this point, we note that as long as the modulation frequency is comparable (within the same order of magnitude) to the collision rate, collisions allow for a redistribution of energy between axes fast enough such that all axes heat at effectively the same rate. This is clear from Fig.~\ref{fig:a92_ad199_Omega90} and Fig.~\ref{fig:a92_ad199_Omega80}, and can be shown true with other values of $\alpha$ and $\Omega$. To this end, we look for a collective transmissibility function.

Using the Greens function, the solution to the driven problem is 
\begin{align}
    \boldsymbol{\xi}(t) &= \boldsymbol{\xi}(0) + \int_{-\infty}^{t} dt' \boldsymbol{G}(t - t') h \boldsymbol{u}_3 \sin[\Omega t'],
    \label{eq:driven_solution}
\end{align} 
where $\boldsymbol{\xi}(0)$ is the solution in the absence of driving. The response $\boldsymbol{\xi}_R(t) \equiv \boldsymbol{\xi}(t) - \boldsymbol{\xi}(0)$ is therefore given by the integral in (\ref{eq:driven_solution}), and is 
\begin{align}
    \boldsymbol{\xi}_R(t) &= \int_{-\infty}^{t} dt' \boldsymbol{G}(t - t') h \boldsymbol{u}_3 \sin[\Omega t'] \\
    &= h \int_{-\infty}^{t} dt' \boldsymbol{G}(t') \boldsymbol{u}_3 \sin[\Omega (t - t')] \\
    &= h \int_{-\infty}^{t} dt' \boldsymbol{G}(t') \boldsymbol{u}_3 \left[ \frac{e^{i\Omega (t - t')} - e^{i\Omega (t - t')}}{2i} \right] \\
\begin{split}
    &= \frac{h}{2i} \bigg[ e^{i\Omega t} \int_{-\infty}^{t} dt' \boldsymbol{G}(t')\boldsymbol{u}_3 e^{-i\Omega t'} \\
    &\quad\quad - e^{-i\Omega t} \int_{-\infty}^{t} dt' \boldsymbol{G}(t')\boldsymbol{u}_3 e^{i\Omega t'}  \bigg] 
\end{split} \\
    &= \frac{h}{2i} \left[ e^{i\Omega t}\tilde{\boldsymbol{G}}(\Omega)\boldsymbol{u}_3 - e^{-i\Omega t}\tilde{\boldsymbol{G}}^*(\Omega)\boldsymbol{u}_3 \right] \\
\begin{split}
    &= \Re\left\{ \tilde{\boldsymbol{G}}(\Omega) \right\} h \boldsymbol{u}_3 \sin(\Omega t) \\
    &\quad\quad + \Im\left\{ \tilde{\boldsymbol{G}}(\Omega) \right\} h \boldsymbol{u}_3 \cos(\Omega t).
\end{split} 
\end{align}
Writing this component-wise and recasting the above expression into the amplitude and phase notation gives
\begin{align}
    [\boldsymbol{\xi}_R (t)]_i = \abs{\left[ \tilde{G}(\Omega)h\boldsymbol{u}_3 \right]_i} \sin(\Omega t + \arg\left[ \tilde{G}(\Omega)h\boldsymbol{u}_3 \right]_i ),
\end{align}
with $[\ldots]_i$ denoting the individual vector elements. The ratio of the amplitude of $\boldsymbol{\xi}_R$ to the drive amplitude $h$ grants the transmissibility function
\begin{align}
    \tau(\alpha, \Omega) = \omega_{z, 0} \sqrt{ \boldsymbol{u}_3^T \left( \boldsymbol{\Phi}_0^2 + \Omega^2 \mathbf{I} \right)^{-1} \boldsymbol{u}_3 }.
\end{align}

\begin{figure}[H]
    \centering
    \includegraphics[width=\textwidth]{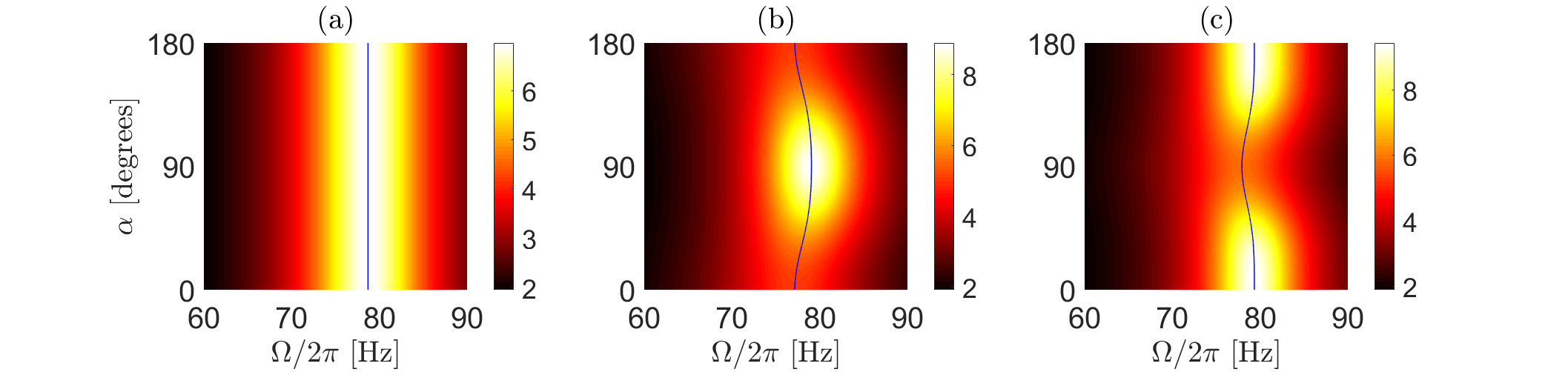}
    \caption{Transmissibility, $\tau(\alpha, \Omega)$: (a) with $a = 109.5 a_0$, $a_d = 0.0$ (isotropic scatterers); (b) with $a = 92.0 a_0$, $a_d = 199.0 a_0$ (dipolar $^{164}$Dy); and (c) with $a = 0.0$, $a_d = 367.2 a_0$ (purely dipolar scatterers). The blue curves in panels (a), (b) and (c) track the peaks of each lineshape as $\alpha$ is varied. Note that increasing the scattering anisotropies ($a_d$ relative to $a$) increases the susceptibility of the gas (larger dipolar character increases the resonant response). }
    \label{fig:transmissibilities}
\end{figure}

To illustrate the effects of scattering anisotropies, we plot $\tau(\alpha, \Omega)$ with 3 different dipolar characteristics: 1) purely isotropic scatterers with no intrinsic dipole, as occurring for instance in a gas with non-polar atomic species; 2) dipolar scatterers with a finite $s$-wave scattering length, representing a gas of $^{164}$Dy atoms as we have considered so far in this paper; 3) purely dipolar scatterers with no $s$-wave scattering length. This range of possibilities can of course be realized by means of the many Fano-Feshbach resonances in the lanthanide species. For the sake of the theoretical comparison presented here,  we somewhat artificially tune the values of $a$ and $a_d$ such that the angular averaged total cross-section,
\begin{align}
    \overline{\sigma} = \frac{1}{2} \int_{-1}^{+1} \sigma(\boldsymbol{p}_{\text{rel}}) d(\cos\eta),
\end{align}
is the same in all three cases, so that each example is in the same collisional regime. The results as presented in Fig.~\ref{fig:transmissibilities} clearly show how the resonant frequencies and energy absorbed by the gas vary with $\alpha$.

\onecolumngrid

\twocolumngrid

Fig.~\ref{fig:transmissibilities}\textcolor{blue}{a} shows the response of the gas in the absence of dipoles, as a heat map of the response $\tau(\alpha,\Omega)$.  This plot establishes that the gas responds near-resonantly at the $\Omega \sim 2\pi \times 80$ Hz frequency expected, slightly shifted to lower frequencies.  It also establishes a characteristic resonance width, the full-width at half maximum (FWHM), $\Delta\Omega = 17.8$ Hz, due to the collisional damping. The resonance is of course independent of $\alpha$ in the absence of dipoles.

Fig.~\ref{fig:transmissibilities}\textcolor{blue}{b} considers the case of native $^{164}$Dy, with $a = 92 a_0$ and $a_d = 199 a_0$. Here the response shows a distinct anisotropy, with the resonance narrowest and least shifted at $\alpha = 90^{\circ}$, while broadening and shifting as $\alpha$ approaches 0 or $180^{\circ}$.  This behavior makes sense, given that the excitation along $z$ drives primarily the $z$ mode depicted in Fig.~\ref{fig:Mode3}.  As explained above, the damping in this mode decreases as $\alpha$ approaches $90^{\circ}$, as collisional excitation in the $y$ direction is not engaged (Fig.~\ref{fig:Mode3}\textcolor{blue}{d}). Therefore, considering the system as a damped, driven oscillator \cite{Taylor}, the resonance is narrower and more strongly peaked when $\alpha = 90^{\circ}$.

\begin{figure}[h]
    \centering
    \includegraphics[width=\columnwidth]{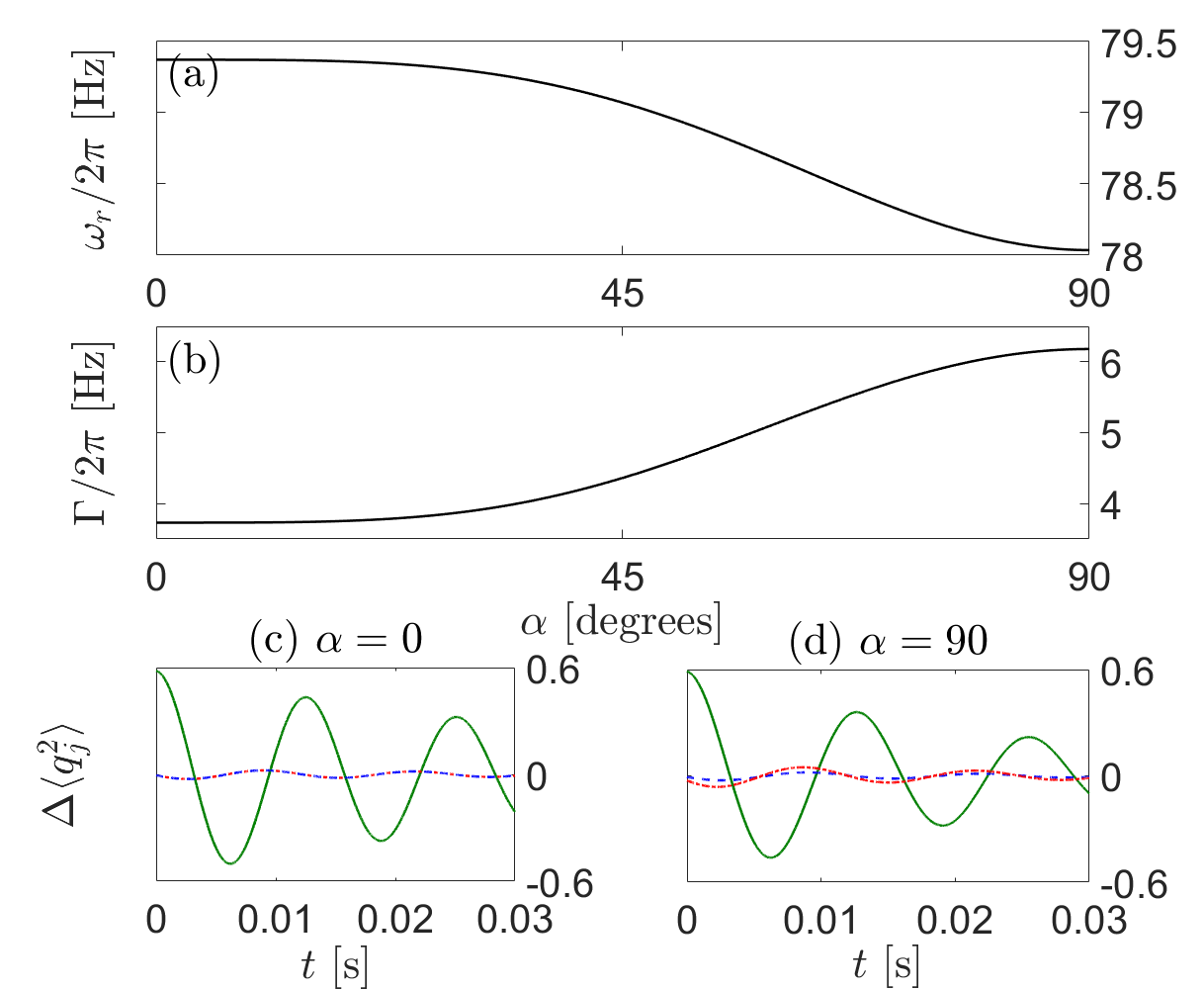}
    \caption{$z$-dominant normal mode solution with $a = 0$, $a_d = 367.2a_0$. Data format, colors and markers follow that in Fig.~\ref{fig:Mode1}. The radial oscillations indicate a preferential scattering into the $y$ axis (larger oscillations in $y$) when $\alpha = 90^{\circ}$. }
    \label{fig:Mode3_a0}
\end{figure}

By contrast, the mode in Fig.~\ref{fig:transmissibilities}\textcolor{blue}{c} shows the opposite trend, with the resonance narrowing in the $\alpha = 0$ and $\alpha = 180^{\circ}$ limits. This is the case of purely dipole scattering, with $a = 0$ and $a_d = 367.2 a_0$. In this a case, the dominant $z$ mode has the characteristics shown in Fig.~\ref{fig:Mode3_a0}.  By contrast to the mode in Fig.~\ref{fig:Mode3}, in this mode the cross section is sufficiently different that the collisional damping rate $\Gamma$ is an increasing function of $\alpha$, thus broadening the response resonance in Fig.~\ref{fig:Mode3_a0}\textcolor{blue}{c} when $\alpha = 90^{\circ}$. The mode for $\alpha = 90^{\circ}$ (Fig.~\ref{fig:Mode3_a0}\textcolor{blue}{d}) illustrates that the excitations in the $x$ and $y$ directions are slightly out of phase, increasing the chance of collisions with momentum components in the $x$-$y$ plane, and therefore increases the overall collision rate $\Gamma$.

\begin{figure}[h]
    \centering
    \includegraphics[width=\columnwidth]{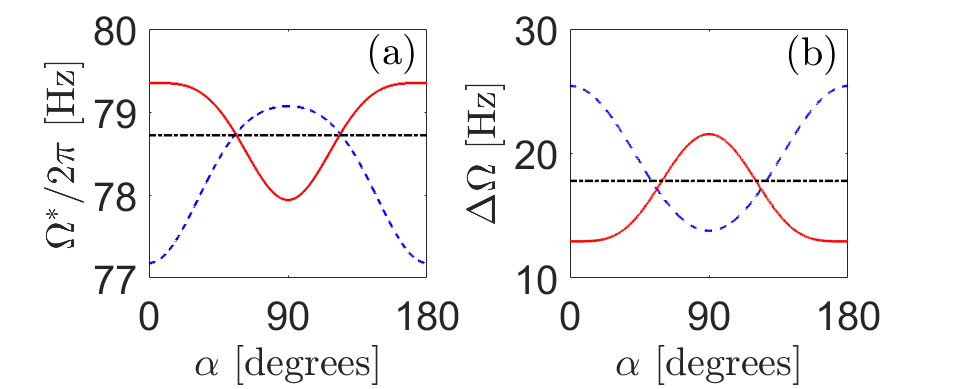}
    \caption{Resonance frequency $\Omega^*$, and FWHM $\Delta\Omega$, of the linear response function as $\alpha$ is varied. (a) $\Omega^*$, as a function of $\alpha$; (b) $\Delta\Omega$, as a function of $\alpha$. Panels (a) and (b) compare between: isotropic scatterers [dot-dashed (black) line]; dipolar scatterers with $a = 92a_0$ [dashed (blue) line]; and purely dipolar scatterers [solid (red) line]. }
    \label{fig:transmissibility_properties}
\end{figure}

To summarize the anisotropy of the line shapes, for each $\alpha$ we extract the resonance frequency $\Omega^*$, and FWHM, $\Delta\Omega$.  The variation in these quantities with $\alpha$ is plotted in Fig.~\ref{fig:transmissibility_properties}. These plots once again showcase the complementary behavior between the gases with and without a scattering length.

\section{\label{sec:MeanField} Discussion of Mean-Field Effects}

Throughout this analysis, we have ignored all phenomena associated to dipolar mean-field effects. In this section, we justify this approximation given the parameters adopted in this work (Table~\ref{fig:a92_ad199_Omega90}). To do so, we consider the total mean-field energy per particle $e_{\text{mf}}$ which can be evaluated analytically for a cylindrical Gaussian with $\langle x^2 \rangle = \langle y^2 \rangle = \langle q_{\perp}^2 \rangle$, \cite{DipolarReview} to give
\begin{align}
    e_{\text{mf}} = -\frac{N}{48\sqrt{\pi^3}} \frac{\mu_0 \mu^2}{\langle q_{\perp}^2 \rangle \sqrt{\langle z^2 \rangle}} h(\rho), 
\end{align}
where
\begin{align}
    h(\rho) = \frac{1 + 2\rho^2}{1 - \rho^2} - \frac{3\rho^2 \text{arctanh}\sqrt{1 - \rho^2}}{(1 - \rho^2)^{3/2}},
\end{align}
with $\rho = \sqrt{\langle q_{\perp}^2 \rangle / \langle z^2 \rangle}$. The function $h(\rho)$ is of order unity, so we just consider the prefactor to get an order of magnitude estimate for $e_{\text{mf}}$. We compare this to the thermal energy per particle $k_B T_0$ with a ratio at thermal equilibrium, which works out to be
\begin{align}
    \frac{1}{k_B T_0} \left( \frac{N}{48\sqrt{\pi^3}} \frac{\mu_0 \mu^2}{\langle q_{\perp}^2 \rangle_0 \sqrt{\langle z^2 \rangle_0}} \right) \approx 9.0 \times 10^{-3}.
\end{align}
This implies that the mean-field effects will indeed be insignificant compared to phenomena associated to kinetic and collisional processes. 

However, in higher density regimes, or for particles with stronger dipole interactions such as polar molecules, mean-field effects would desirably be included into the model. Physics associated to such effects could present a wide variety of interesting dynamical observations. We defer further discussions on this to future publications.

\section{\label{sec:conclusions} Conclusions}

The nonequilibrium thermodynamics of the ultracold, dipolar gas depends strongly on the anisotropy of the differential scattering  of the dipolar constituents of the gas.  Thus anisotropy plays a significant role even for ultracold gases that are not quantum degenerate.  To study the macroscopic dynamics, we have derived closed-form expressions that constitute the Enskog equations up to linear-order from thermal equilibrium. At short times and for weak driving, these expressions are in excellent agreement with direct Monte Carlo simulations. This suggests that they provide a quantitative means for us to understand such dipolar systems when probed perturbatively. The extension of the Enskog formalism to fermionic species is of course possible, as the differential cross section is known, and this will be a subject of future investigations.  

The resulting normal modes illustrate the strong dependence of the parametric heating rate of the gas on both the $s$-wave scattering length and the size and orientation of the atomic dipoles.  Therefore, as a function of these experimentally controllable parameters, the gas becomes a working fluid whose response to perturbation can be manipulated, which may lead to further investigations and applications down the line.  Strikingly, the normal mode analysis also identifies melting modes, anisotropic distortions of the gas that equilibrate without exciting oscillations, even in a harmonic trap.  The significance of these modes and prospects for their observation, will be considered in future work.

\begin{acknowledgments}

This material is based upon work supported by the National Science Foundation under Grant Number PHY 1734006 and Grant Number PHY 1806971.

\end{acknowledgments}

\appendix

\section{\label{app:DipolarCollInt} The Dipolar Collision Integral}

An explicit derivation of the collision integrals in the Enskog formulation (CIE), $\mathcal{C}[\Delta p_j^2]$, is presented here. To prevent an over cluttering of this manuscript, several intermediate expressions are omitted. Many of these are exceedingly long and not particularly illuminating, involving integrals over polynomial functions that can be evaluated with most modern symbolic software. 

The starting point to evaluate the CIE is the assertion of a Gaussian ansatz
\begin{subequations}
\label{eq:Gaussian_ansatz}
\begin{align}
    & f(\boldsymbol{q}, \boldsymbol{p}) = c(\boldsymbol{p}) n(\boldsymbol{q});  \\
    & c(\boldsymbol{p}) \equiv \prod_{j} \frac{1}{\sqrt{2\pi \langle p_j^2 \rangle}} \exp\left( -\frac{p_j^2}{2\langle p_j^2 \rangle} \right), \\
    & n(\boldsymbol{q}) \equiv N \prod_{j} \frac{1}{\sqrt{2\pi \langle q_j^2 \rangle}} \exp\left( -\frac{q_j^2}{2\langle q_j^2 \rangle} \right),
\end{align}
\end{subequations}
with subscripts $j \in \{x, y, z\}$. If we now recast the momenta into center of mass coordinates, we get a decomposition into a center of mass component $\boldsymbol{P}$ and a relative component $\boldsymbol{p}_r$. This grants the reformulation of the CIE from Eq.~(\ref{eq:Enskog_collision_integral}) to
\begin{align}
    \mathcal{C}[\Delta \chi] = \int d^3q \frac{n^2(\boldsymbol{q})}{N} & \int \frac{d^3p_r}{2m} p_r c_r(\boldsymbol{p}_r) \nonumber\\
    \times & \int d\Omega_{p'} \frac{d\sigma}{d\Omega_{p'}} \Delta\chi
\end{align}
where $c_r(\boldsymbol{p}_r)$ takes the same form of $c(\boldsymbol{p})$ but with the replacement $\boldsymbol{p} \rightarrow \boldsymbol{p}_r$ and all factors of 2 converted to 4. As mentioned, only the $\chi = p_j^2$ terms are non-vanishing, so the integrals are separable (i.e. can be evaluated separately) in position and momentum variables
\begin{align}
    \mathcal{C}[\Delta p_j^2 ] = & \left[ \int d^3q \frac{n^2(\boldsymbol{q})}{N} \right] \nonumber\\
    \times & \left[ \int \frac{d^3p_r}{2m} p_r c_r(\boldsymbol{p}_r) \int d\Omega_{p'} \frac{d\sigma}{d\Omega_{p'}} \Delta p_j^2 \right]. \label{eq:CIE_recast}
\end{align}
First evaluating the integral over $d^3 q$ gives
\begin{align}
    \mathcal{I}_q \equiv \frac{1}{N} \int d^3q \: n^2(\boldsymbol{q}) = \frac{N}{8 m \sqrt{\pi^{3} \langle x^2\rangle \langle y^2\rangle \langle z^2\rangle}},
\end{align}
which in the linearization, adopts the variance values at thermal equilibrium. This leaves the integrals over post and pre-collision momenta. 

Evaluating the momentum integrals in Eq.~(\ref{eq:CIE_recast}) is a difficult task for dipoles, predominantly due to the fact that the differential cross-section is anisotropic, preventing its factorization out of the integral. This requires the coordinate frames for integration to be consistent and carefully handled. In our approach, we define 2 relevant coordinate frames:
\begin{enumerate}
    \item \textit{the laboratory-frame} (lf), defined with respect to the dipole alignment axis such that
    \begin{align}
        \hat{\varepsilon} = \big[ \sin\alpha, \:\: 0, \:\: \cos\alpha \big]^T;
    \end{align}
    
    \item \textit{the collision-frame} (cf), defined by the relative momenta of colliding pairs of atoms with
    \begin{align}
        \hat{z}_{\text{cf}} = \hat{\boldsymbol{p}}_r. 
    \end{align}
\end{enumerate}

It is necessary to perform integrals over both the lf coordinates $\{p_r, \theta, \phi\}$ in which $\alpha$ is defined, and cf coordinates $\{\theta_{\text{cf}}, \phi_{\text{cf}}\}$ that defines the post-collision relative momentum (subscript cf is used instead of primes to be unambiguous about the frame). As such, a transformation that relates these 2 sets of variables is necessary and constructed using the method of direction cosines
\begin{align}
    R(\text{cf}\:\rightarrow\:\text{l})  = \begin{bmatrix}
        \hat{x}\cdot\hat{x}_{\text{cf}} & \hat{x}\cdot\hat{y}_{\text{cf}} & \hat{x}\cdot\hat{z}_{\text{cf}}  \\
        \hat{y}\cdot\hat{x}_{\text{cf}} & \hat{y}\cdot\hat{y}_{\text{cf}} & \hat{y}\cdot\hat{z}_{\text{cf}} \\
        \hat{z}\cdot\hat{x}_{cf} & \hat{z}\cdot\hat{y}_{\text{cf}} & \hat{z}\cdot\hat{z}_{\text{cf}}
    \end{bmatrix}.
\end{align}

With this, the differential cross-section can be obtained in the cf with Eq.~(\ref{eq:scattering_amplitude}), for which the unit vectors in the cf are given as
\begin{align}
    & \hat{k} = \begin{bmatrix}
        0 \\ 0 \\ 1
    \end{bmatrix}, \quad \hat{k}' = \begin{bmatrix}
        \sin\theta_{\text{cf}} \cos\phi_{\text{cf}} \\ 
        \sin\theta_{\text{cf}} \sin\phi_{\text{cf}} \\ 
        \cos\theta_{\text{cf}}
    \end{bmatrix}, \\
    & \hat{\varepsilon} = \begin{bmatrix}
        \sin\alpha \sin\phi \\
        \sin\alpha \cos\theta \cos\phi - \cos\alpha \sin\theta \\
        \sin\alpha \sin\theta \cos\phi + \cos\alpha \cos\theta
    \end{bmatrix}.
\end{align}

These result in the dot product terms
\begin{subequations}
\begin{align}
    & \hat{k}\cdot\hat{k}' = \cos\theta_{\text{cf}}, \\
    & \hat{k}\cdot\hat{\varepsilon} = \sin\alpha \sin\theta \cos\phi + \cos\alpha \cos\theta, \\
    & \hat{k}'\cdot\hat{\varepsilon} = \sin\theta_{\text{cf}} \cos\phi_{\text{cf}}\sin\alpha \sin\phi \\
    &\quad\quad + \sin\theta_{\text{cf}} \sin\phi_{\text{cf}} \left( \sin\alpha \cos\theta \cos\phi - \cos\alpha \sin\theta \right) \nonumber \\
    &\quad\quad + \cos\theta_{\text{cf}} \left( \sin\alpha \sin\theta \cos\phi + \cos\alpha \cos\theta \right). \nonumber
\end{align}
\end{subequations}

Plugging these into Eq.~(\ref{eq:scattering_amplitude}) and taking its absolute-square gives the differential cross-section, which can be factorized into terms of various orders in $a_d$ as functions of $\alpha$,
\begin{align}
    \frac{d\sigma_B}{d\Omega_{p'}}(\alpha) = \frac{d\sigma_B^{(0)}}{d\Omega_{p'}}(\alpha) + \frac{d\sigma_B^{(1)}}{d\Omega_{p'}}(\alpha) + \frac{d\sigma_B^{(2)}}{d\Omega_{p'}}(\alpha),
\end{align}
where the superscripts on each term indicate the order of $a_d$ dependence. This integral is then done term-by-term over $d\Omega_{p'}$ written as
\begin{align}
    \mathcal{I}_p^{\text{(cf)}}(p_r, \theta, \phi) \equiv \int d\Omega_{p'} \left[ \frac{d\sigma_B^{(0)}}{d\Omega_{p'}} + \frac{d\sigma_B^{(1)}}{d\Omega_{p'}} + \frac{d\sigma_B^{(2)}}{d\Omega_{p'}} \right] \Delta p_j^2.
\end{align}

Now comes the integral over laboratory-frame coordinates. To do this, a Taylor expansion of $c_r(\boldsymbol{p}_r)$ is first done up to first-order around thermal equilibrium 
\begin{align}
    c_{r}(\boldsymbol{p}_r) \approx\: c_r^{\text{eq}}(\boldsymbol{p}_r) \bigg[ & \delta p_x \left( \frac{p_{r}^2 \sin^2\theta \cos^2\phi }{4\langle p_z^2 \rangle_{0}} - \frac{1}{2}\right) \nonumber \\
    + & \delta p_y \left( \frac{p_{r}^2 \sin^2\theta \sin^2\phi }{4\langle p_z^2 \rangle_{0}} - \frac{1}{2}\right) \nonumber \\
    + & \delta p_z \left( \frac{p_{r}^2 \cos^2\theta }{4\langle p_z^2 \rangle_{0}} - \frac{1}{2}\right) + 1\bigg], 
\end{align}
where 
\begin{align}
    & \delta_{p_j} \equiv \frac{\langle p_j^2 \rangle}{\langle p_z^2 \rangle_{0}} - 1 \\
    & c_r^{\text{eq}}(\boldsymbol{p}_r) \equiv \dfrac{ 1 }{ \sqrt{(4\pi \langle p_z^2 \rangle_{0} )^{3}} } \exp\left( -\frac{ p_{r}^2 }{4\langle p_z^2 \rangle_{0}} \right),
\end{align}
with $c_r^{\text{eq}}(\boldsymbol{p}_r)$ being the equilibrium distribution of relative momenta. It is noted that all terms in $c_r(\boldsymbol{p}_r)$ with constant coefficients multiplying $c_r^{\text{eq}}(\boldsymbol{p}_r)$ are trivial since the collision integral vanishes at thermal equilibrium. Putting all this together gives
\begin{align}
    \mathcal{C}[\Delta p_j^2 ](\alpha) = \mathcal{I}_q \int \frac{d^3p_r}{2m} p_r c_r(\boldsymbol{p}_r) \mathcal{I}_p^{\text{(cf)}}(p_r, \theta, \phi),
\end{align}
which when evaluated, leads to Eqs.~(\ref{eq:EnskogColl}).

\nocite{*}

\bibliography{main.bib} 

\end{document}